\documentclass[12pt]{iopart}
\usepackage{iopams}
\usepackage{graphicx}
\begin{document}

\letter{On the detectability of gravitational waves background
produced by gamma ray bursts}

\author{Giulio Auriemma}

\address{Universit\`{a} degli Studi della Basilicata, Potenza, Italy}
\begin{abstract}
In this paper we discuss a new strategy for the detection of
gravitational radiation likely emitted by cosmological gamma ray
burst. Robust and conservative estimates lead to the conclusion
that the uncorrelated superimposition of bursts of gravitational
waves can be detected by interferometric detectors like VIRGO or
LIGO. The expected signal is predicted to carry two very
distinctive signatures: the cosmological dipole anisotropy and a
characteristic time scale in the auto correlation spectrum, which
might be exploited, perhaps with \emph{ad hoc} modifications
and/or upgrading of the planned experiments, to confirm the
non-instrumental origin of the signal.
\end{abstract}

%Uncomment for PACS numbers title message
\pacs{04.80.Nn, 95.55.Ym, 98.70.Rz, 98.70.Vc}

% Uncomment for Submitted to journal title message
%\submitto{\JPG}

% Comment out if separate title page not required
%\maketitle
\section{Introduction}\label{sect:intro}
Gamma Ray Bursts (GRB) are short and intense flashes of e.m.
radiation in the keV-MeV range, accidentally discovered more then
30 years ago \cite{review}. The observation with high spatial
resolution by the Italian-Dutch satellite Beppo-SAX \cite{Costa97}
of the X-ray afterglow, has allowed the optical identification of
some of the burst, which has given the evidence of the
cosmological origin of the GRB's. It is now evident that for few
seconds, at least twice a day, unknown sources become so bright to
over shine all the visible Universe \cite{Piran-rev}. From the
observed $\gamma$-ray fluence and the red shift of the host
galaxy, it is estimated \cite{Jimenez-Band-Piran01} that the
average buster emits isotropically $\langle
E_{\gamma\,iso}\rangle=\left(1.3^{+1.2}_{-1}\right)\times
10^{53}\; \mathrm{erg}$ (We will assume \cite{Fukugita-Hogan00}
here and in the rest of the paper a flat Universe with
$H_0=65\;\mathrm{km}\;\mathrm{s}^{-1}\,\mathrm{Mpc}^{-1}$,
$\Omega_M=0.3$ and $\Omega_V=0.7$). The dispersion around this
average is consistent with a log-normal distribution having
logarithmic width $\sigma_{\gamma\, iso}=1.7 ^{+0.4}_{-0.3}$
(about a factor 50). This energy corresponds in the average to
$\approx 0.1\,M_{\odot}\,c^2$, but arrives in the extreme case of
GRB 990123 \cite{Kulkarni99} to be $\approx 1.4\,M_{\odot}\,c^2$,
which poses serious problems for understanding what could be the
``central engine'' powering these events. In the currently
accepted ``fireball'' model \cite{review} the $\gamma$--rays are
emitted by electron/positron pairs, accelerated by internal
relativistic shocks, which radiates via synchrotron and/or
synchro-Compton process. In order to be optically thin, the
fireball should expand with ultra--relativistic velocity (bulk
Lorentz factor $\Gamma\approx 100$). Even if the conversion of the
kinetic energy into $\gamma$ radiation is estimated
\cite{Guetta-Spada-Waxman01} to be very efficient
($\eta_{\gamma}\approx 0.2$), the formation of the relativistic
wind with a kinetic energy $\approx E_{\gamma\,
iso}/\eta_{\gamma}$ could be problematic.

The energy output of the central engine of GRB can be largely
reduced if the $\gamma$-ray emission is beamed with a small
opening angle. This possibility is supported by the observation of
an achromatic break in the time evolution of the afterglow
luminosity, occurring when the bulk Lorentz factor becomes
$\Gamma<\theta _{jet}^{-1}$. In a recent paper \cite{Frail01} this
hypothesis is applied to estimate the opening angle of the 10
GRB's with observed afterglows, with the interesting result that
the average opening angle in the sample is $\langle\theta
_{jet}\rangle\simeq 4^\circ$. Furthermore, it appears from these
data that the fluence of the GRB in the sample is correlated with
the opening angle, in a way that justifies the large spread of the
apparent fluence as entirely due to the spread of the opening
angles. The intrinsic energy emitted by GRB in $\gamma$-rays, if
estimated as
\begin{equation}\label{eq:intro1}
  E_{\gamma\,jet}=f_{jet}\,F_\gamma\,d_L^2\,(1+z)^{-1}
\end{equation}
where $f_{jet}=\frac{1}{2}(1-\cos\theta_{jet})$ is the beaming
factor \cite{Frail01}, clusters around a value of $\langle
E_{\gamma\,jet}\rangle=5\times10^{50}\;\mathrm{erg}$ with a
logarithmic width $\sigma_{\gamma\,jet}\approx 0.3$ that is about
a factor 25 smaller than the one of the $E_{\gamma\, iso}$. If
this is true the kinetic energy would be of the order of
$E_{GRB}\approx 10^{-3}\,M_{\odot}\,c^2$, about constant for all
the bursts. The scenario emerging from this study is particularly
appealing because one could assume that the central engine of the
GRB produces always a similar amount of energy, while the
complexity of the energy transfer from the central engine to the
relativistic jet causes the wide range of opening angle, and
consequently the wide range of apparent fluence that is observed.

This scenario implies also that the true rate of bursts
$R_{jet}\approx \left\langle f_{jet}^{-1}\right\rangle\,R_{iso}$,
is enhanced by a factor $\left\langle
f_{jet}^{-1}\right\rangle\approx 500$ respect to the observed
rate. However clearly the total energy output per unit volume and
unit time injected in the universe by GRB explosion
$\dot\epsilon_{\gamma}=R_{jet}\,E_{\gamma\,jet}=R_{iso}\,E_{\gamma\,
iso}$ will be the same of the isotropic case.

In the next Sect.\ \ref{sect:single}\ of this paper we discuss the
implications of the possible beamed emission on the detectability
of gravitational waves (GW) likely emitted by the collapse event
that originates the GRB. It is to be expected that if the e.m.
energy of the $\gamma$-ray burst is not extraordinary (about the
same order of magnitude of the total output of a Type Ib SN), the
energy output in gravitational waves should also be modest. But,
as we observed above, in the latter case the number of bursts
should be increased correspondingly. This suggests a new strategy
for looking to gravitational radiation from GRB. As we will show
in Sect.\ \ref{sect:back}, even if the GW's produced by each GRB
are below the detection threshold, integrating over one year one
should find an excess of noise, due to the uncorrelated
superimposition of many GW pulse trains. In Sect.\
\ref{sect:signa} we discuss two very distinctive signature that a
genuine cosmological stochastic signal should carry. Those
intrinsic signatures, perhaps detectable with modifications and/or
upgrading of the planned experiments, could be of great help to
check the non-instrumental origin of the stochastic signal.
Finally in Sect.\ \ref{sect:concl} we summarize the results of
this investigation.
\section{Gravitational wave signal from GRB}\label{sect:single}
The energy required to power the $\gamma$-ray burst, estimated by
Eq.\ (\ref{eq:intro1}), is of the order of $\approx
10^{-3}\,M_\odot\,c^2$, not overwhelming but always much larger
then the total nuclear binding energy of few-$M_\odot$'s stars.
Therefore it requires in any case an energy release that is
compatible only with the collapse of a several solar masses
object. A natural assumption would be that the GRB are similar to
ordinary supernovas, but perhaps the difference is in the final
result of the collapse. Like for the ``failed supernova'' model
\cite{Woosley93} the collapse of a massive star to a black hole
surrounded by a dense rotating torus of material that might result
in a relativistic jet. The energy irradiated in this case as
gravitational waves can be estimated \cite{Stark-Piran85} as
$E_{GW}\approx \epsilon_{GW}\, M_{BH}\,c^2$ where $M_{BH}$ is the
mass of the black hole and $\epsilon_{GW}\lesssim 10^{-4}$ is an
efficiency parameter. As we observed above the bolometric energy
released in the gamma ray burst is already of the order of
$E_{\gamma\,jet}\gtrsim 10^{-4}\,M_{\odot}\,c^2$ and
$\eta_{\gamma}\approx 0.2$, then we could expect that the energy
provided by the central engine could be $E_{GRB}\approx
\epsilon_{GRB}\,M_{BH}\,c^2$ where $\epsilon_{GRB}$ is the
fraction of the gravitational energy converted into kinetic energy
of the ejecta subsequently radiated as $\gamma$-rays.

It could be possible under certain conditions \cite{van Putten01}
that $\epsilon_{GRB}\approx \epsilon_{GW}$, but perhaps we should
adopt the more conservative view that $\epsilon_{GW} \lesssim
\epsilon_{GRB}$ even if it is unlikely that
$\epsilon_{GW}\ll\epsilon_{GRB}$. Our ignorance of the actual
mechanism that could transfer the energy from the accretion torus
to the jet does not allow a solid prediction of the energetic of
the central engine of the GRB. Nevertheless we can set a lower
limit to this energy in the form $E_{GRB}\gtrsim
E_{\gamma\,jet}/\eta_{\gamma}$. We can parameterize the production
of GW introducing a phenomenological $\eta_{GW}\lesssim 1$ putting
$E_{GW}=\eta_{GW}\,E_{\gamma\,jet}/\eta_{\gamma}$ and verifying
the sensitivity of the GW detectors to the actual value of
$\eta_{GW}$.

The energy flux (\emph{viz.} energy per unit surface and unit
time) of GW produced by a burst of intrinsic luminosity (in
gravitational waves) $L_{GW}$ at a red shift $z$ is given by
\cite{Kolb-Turner00}
\begin{equation}
    F^b_{GW}=\frac{(1+z)\,L_{GW}}{4\pi\,d_{L}^2}
    \label{eq:single1}
\end{equation}
where $d_{L}$ is the luminosity distance. Assuming a typical time
scale for the emission $\Delta t$ we can estimate this flux at
Earth for a typical red shift $z=1$
$$F^b_{GW}(z=1)\approx
  4\times 10^{-7}\,
  \left(\frac{f_{jet}\,E_{\gamma iso}/\eta_\gamma}{2.5\times10^{51}\;\mathrm{erg}}\right)\,
  \left(\frac{\eta_{GW}}{0.01}\right)\,\left(\frac{10\;\mathrm{ms}}{\Delta t}\right)
  \quad\mathrm{erg}\,\mathrm{cm} ^{-2}\,\mathrm{s} ^{-1}\,\,\mathrm{sr} ^{-1}$$
   In order to convert the flux of Eq.\ (\ref{eq:single1}) into an
adimensional amplitude we can use the classical formula
\cite{Shapiro-Teukolsky83}:
\begin{equation}
    F_{b}=\frac{c^{3}}{16\pi G}\,\left\langle \dot
    h^{2}_{+}+\dot h^{2}_{\times}\right\rangle
    \label{eq:single2}
\end{equation}
where the average is taken over several wavelengths. The amplitude
of the signal produced depends from the direction and the beam
pattern of the detector. In the best case we have integrating over
time and applying the Parseval' s theorem
\begin{equation}
    \int_{-\infty}^{+\infty} \omega^2\,\tilde h_{b}^{2}(\omega)\,d\omega=
    \frac{16\pi G}{c^{3}}\frac{(1+z)\,E_{GW}}{4\pi\,d_{L}^2}
    \label{eq:single3}
\end{equation}
where $\tilde h_{b}^{2}(\omega)$ will be the Rayleigh power of the
signal as a function of the frequency $\omega$. In order to
estimate the order of magnitude of the amplitude of the GW signal
we do not need a detailed shape of the spectral power density of
the signal, but only the knowledge of the firsts two moments of
the distribution $\bar\omega$ and $\Delta\omega$. In fact we can
recast Eq.\ (\ref{eq:single3}) in the form
$$\left(\bar
\omega^{2}+\Delta\omega^{2}\right)\,\int_{-\infty}^{+\infty}
\tilde h_{b}^{2}(\omega)\,d\omega= \frac{16\pi
G}{c^{3}}\frac{(1+z)\,E_{GW}}{4\pi\,d_{L}^2}$$ It is remarkable
that rather natural physical assumption on the first and second
moment of the unknown Rayleigh power distribution $\tilde
h_{b}^{2}(\omega)$ can be made. In fact we can assume that the
first moment will be $\bar \omega\approx 2\pi\,c/r_{S}$ where
$r_{S}$ will be the Schwarzschild radius of the collapsed object
The second moment will be the r.m.s. bandwidth of a wave packet
that can be estimated $\Delta\omega\approx 2\pi/\Delta t$ where
$\Delta t$ is the duration of the emission in the comoving frame.
In this case we have, if $c\,\Delta t\gg r_{S}$, for the peak
amplitude
\begin{equation}\label{eq:single4}
    \tilde h_b^{peak}  \approx  \frac{r_{S}}{d_{L}}\,
    \sqrt{\frac{G}{2\pi^{3}\,c^{5}}\,E_{GW}\,\Delta t}
\end{equation}
The expected amplitude for typical values is
$$d_L\,\tilde h_b^{peak}\approx
  10^{-27}\,
  \left(\frac{r_S}{5\;\mathrm{km}}\right)\,
  \left(\frac{E_{GRB}}{2.5\times10^{51}\;\mathrm{erg}}\right)^{\frac{1}{2}}\,
  \left(\frac{\eta_{GW}}{0.01}\right)^{\frac{1}{2}}\,
  \left(\frac{\Delta t}{10\;\mathrm{ms}}\right)^{\frac{1}{2}}
  \;\mathrm{Gpc}/\sqrt{\mathrm{Hz}}$$
It is clear from this estimate that the probability of detecting a
single burst is very low, unless $\eta_{GW}\gg1$. However even if
the individual burst could not be detected, it is to be remarked
that in case of beaming the rate of explosion would be very large
($\approx 1500$ per day), therefore the stochastic accumulation of
signal integrated over a long time could emerge from the noise.
\section{Cosmological background}\label{sect:back}
The energy flux of GW produced at Earth from a cosmological
distribution of sources is given by
\begin{equation}
    F^{diff}_{GW}= \int_{0}^{z_{max}}\, R(z)\,E_{GW}\,\frac{dz}{(1+z)\, H(z)}
    \label{eq:back1}
\end{equation}
where $R(z)$ is the comoving rate of bursts per unit volume
exploding at red shift $z$, $E_{GW}$ the average energy emitted in
gravitational waves by each source that we have estimated in
\S\ref{sect:single} and $H(z)=\dot R/R$ is the Hubble expansion
rate \cite{Kolb-Turner00}. As obvious, the lower apparent
brightness of distant burst is compensated by the increase of
volume up to a red shift 1-2, like in the classical Olbert' s
paradox. It is also worth noticing that the flux will be the same
for beamed or isotropic sources. The rate $R(z)$ can be obtained
from the Log N-Log P distribution \cite{Kommers00,Wijers98}, even
if its value depends from the assumption made on the evolution of
the burst rate in the recent past. In practice the energy flux of
the present stochastic background of GW is dominated by the rate
of explosions at $z\gtrsim 1$. If the GRB are produced by a final
catastrophic collapse event of an evolved massive star, one
expects that the large scale distribution of GRB should reflect
the star formation rate. The latter is known
\cite{Lilly96,Madau96} to increase by at least a factor ten from
$z=0$ to $z\approx 2$ to flatten in the range $2\le z\le 3$ and to
decrease exponentially for $z>3$, reaching a value similar to the
present for $z\approx 6$. The fit obtained with this evolution
\cite{Wijers98} is $R_{iso}(z=0)=0.14\pm
0.02\;\mathrm{Gpc}^{-3}\,\mathrm{y}^{-1}$. On the other side if no
evolution is assumed the constant rate is about one order of
magnitude larger. But it is worth noticing that due to the
constraint of having about 1000 bursts per year the only
difference between the two cases is that in the first case the
average distance of the bursts is slightly larger, due to the fast
rise of the rate of explosion with red shift. In fact from Eq.\
(\ref{eq:back1}) we estimate for constant rate
$$F^{diff}_{GW}\approx
  10^{-10}\,
 \left( \frac{E_{iso}/\eta_\gamma}{6.5\times10^{53}\;\mathrm{erg}}\right)\,
  \left(\frac{\eta_{GW}}{0.01}\right)\,
  \left(\frac{R_{iso}}{1\;\mathrm{Gpc}^{-3}\,\mathrm{y}
  ^{-1}}\right)
  \quad\mathrm{erg}\,\mathrm{cm} ^{-2}\,\mathrm{s} ^{-1}\,\,\mathrm{sr} ^{-1}$$
  while assuming evolution of the rate in the recent past (see
  below) the predicted flux is practically the same. Comparing with
  Eq.\ (\ref{eq:single1}) we observe that the diffuse flux is
  consistent with the order of magnitude estimate $$F^{diff}_{GW}/F^b_{GW}\approx\left\langle
f_{jet}^{-1}\right\rangle\times
1,000\;\mathrm{bursts}/1\;\mathrm{year}\times
   \Delta t$$

 Applying again Eq.\
(\ref{eq:single2}) we have:
\begin{equation}
    \frac{c^{3}}{16\pi G}\,\left\langle \dot
    h^{2}_{+}+\dot h^{2}_{\times}\right\rangle=
    \int_{0}^{z_{max}}\, R(z)\,E_{GW}\,\frac{dz}{(1+z)\, H(z)}
    \label{eq:back2}
\end{equation}
On the L.H.S. of this equation we have the amplitude of the wave that invests
at a certain instant the detector. We have seen in the
previous section that each of this burst will not have a detectable
intensity, but if we average over an observation time  $T$ long compared to
the GW burst duration but short compared to Hubble time scale (typically
one year) we have a signal
\begin{equation}
    \frac{1}{T}\int_{-\infty}^{+\infty}\,\omega^{2}\, \tilde
    h^{2}(\omega)\,d\omega=
    \int_{0}^{z_{max}}\, R(z)\,E_{GW}\,\frac{dz}{(1+z)\, H(z)}
    \label{eq:back3}
\end{equation}
that will be detectable if the power spectral density is greater
then the power spectral of the noise, averaged over the same
observation time.
\begin{figure}
\begin{center}
\includegraphics{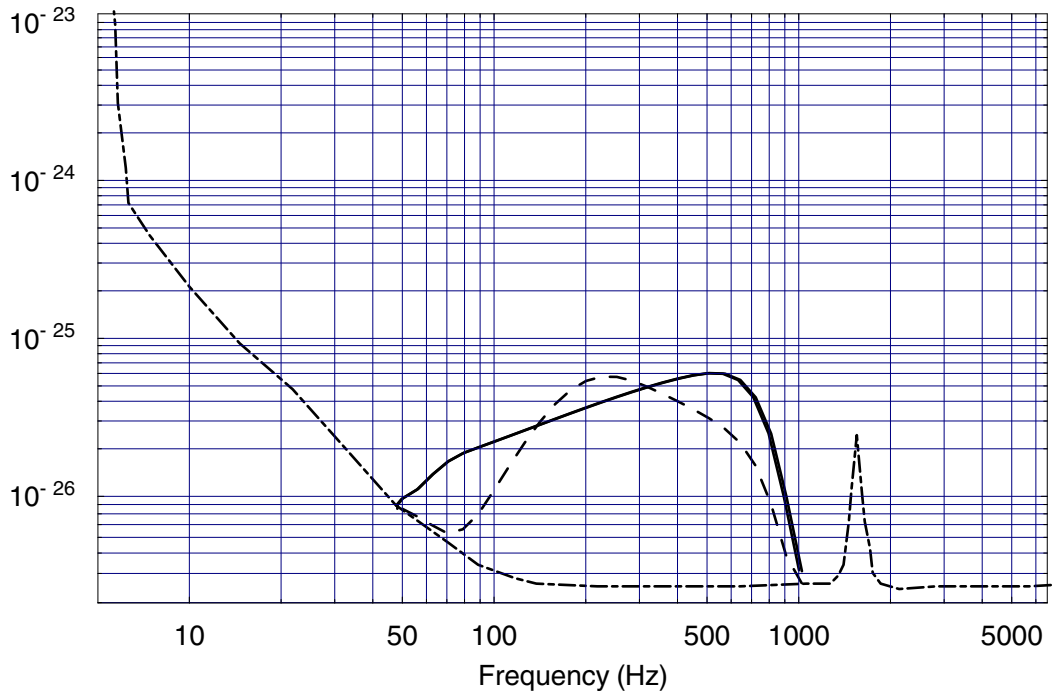}
\end{center}
\caption{\label{fig2} Stochastic signal integrated over one year
from cosmological GRB ($\eta_{GW}=0.01$). In this plot the solid
line corresponds to a flat rate, while the dashed one to a rate of
bursts evolving like the QSO rate (see text). The dot-dashed line
is the noise estimated in the VIRGO proposal \cite{VIRGO},
averaged over one year.}
\end{figure}
The uncorrelated superimposition of bursts of gravitational waves
will be well approximated, for the central limit theorem, by the
 superimposition of red shifted gaussian distributions. Therefore the power
 spectral density of the signal can be estimated by the integral
\begin{equation}
     \langle\tilde h^{2}(\omega)\rangle_{T}\approx
     \frac{G}{\pi^{2}\,c^{5}}\,\int_{0}^{z_{max}}
     \, R(z)\frac{E_{GW}\,r_{S}^{2}}{\sqrt{2\pi}\,\Delta\omega}\,
    e^{-\frac{\left[(1+z)\omega-\omega_{0}\right]^2}{2\Delta\omega^{2}}}\,
    \frac{dz}{(1+z)\, H(z)}
    \label{eq:back4}
\end{equation}
where the normalization has been obtained from Eq.\
(\ref{eq:back3}). We have reported in \Fref{fig2}, which is the
central result of this paper, the Rayleigh power of the stochastic
signal from the cosmological background for $\eta_{GW}=0.01$. The
dashed curve in\ \Fref{fig2} represents the power expected
allowing for an evolution of the rate $R(z)$ with the red shift
similar to the luminosity density evolution of QSO's
\cite{Waxmann-Bahcall98}. This evolution is assumed to be
\cite{Boyle-Terlevich98} $R(z) = (1 + z)^\alpha$ with
$\alpha\simeq 3$ at low red shift, $z < 1.9$, $\alpha = 0$ for
$1.9 < z < 2.7$ \cite{Hewett93}, and an exponential decay at $z
> 2.7$ \cite{Schmidt95}, similar to that describing the evolution of star formation rate
that we have discussed above. The solid curve represents, on the
contrary, the power expected if the rate of the GRB is assumed to
be constant. For comparison we have also reported the noise
expected in the VIRGO experiment \cite{VIRGO} averaged over one
year of integration. This comparison shows that the cosmological
signal should be detectable at the same level in both cases,
because the overall normalization is constrained by the observed
rate by BATSE of $\approx 1000$ bursts per year.

\section{Signatures of the genuine signal}\label{sect:signa}
We expect for the stochastic signal produced by cosmological
sources some very clear signatures showing its origin. We will not
discuss in the following whether or not those signatures will be
detectable by the two planned detectors VIRGO and LIGO, but we
expect that an \emph{ad hoc} modification of those experiments
could be designed in order to extract all the possible
informations from the signal. The more evident of those signatures
should be the dipole anisotropy, as observed for the cosmic
microwave background (CMB). The dipole anisotropy of the CMB is
interpreted as the result of Doppler shift caused by the solar
system motion relative to the isotropic radiation field. This
motion is confirmed by measurements of the apparent velocity of
local galaxies \cite{Dekel99}. The motion of the observer
(receiver) with velocity $\beta=v/c$ relative to a source of
gravitational waves with frequency $\omega_0$ produces a shift in
the observed frequency $\omega'_0$ given by the formula
\cite{Smoot-Scott00}:
\begin{equation}\label{eq:signa1}
  \omega'_0=\omega_0\,\frac{(1-\beta^2)^\frac{1}{2}}{1-\beta\cos\theta}\approx
  \omega_0\,[1-\beta\cos\theta+\mathcal{O}(\beta^2)]
\end{equation}
where the velocity for the solar-system barycentre is
\cite{Kogut93,Lineweaver96} $\beta = 0.001237 \pm 0.000002$ or $v
= 371 \pm 0.5\;\mathrm{km}\,\mathrm{s}^{-1}$ and $\theta$ is the
angle formed with the direction of equatorial coordinates
$(\alpha, \delta) = (11.20^\mathrm{h} \pm
0.01^\mathrm{h},-7.22^\circ \pm 0.08^\circ)$.
\begin{figure}
\begin{center}
\includegraphics{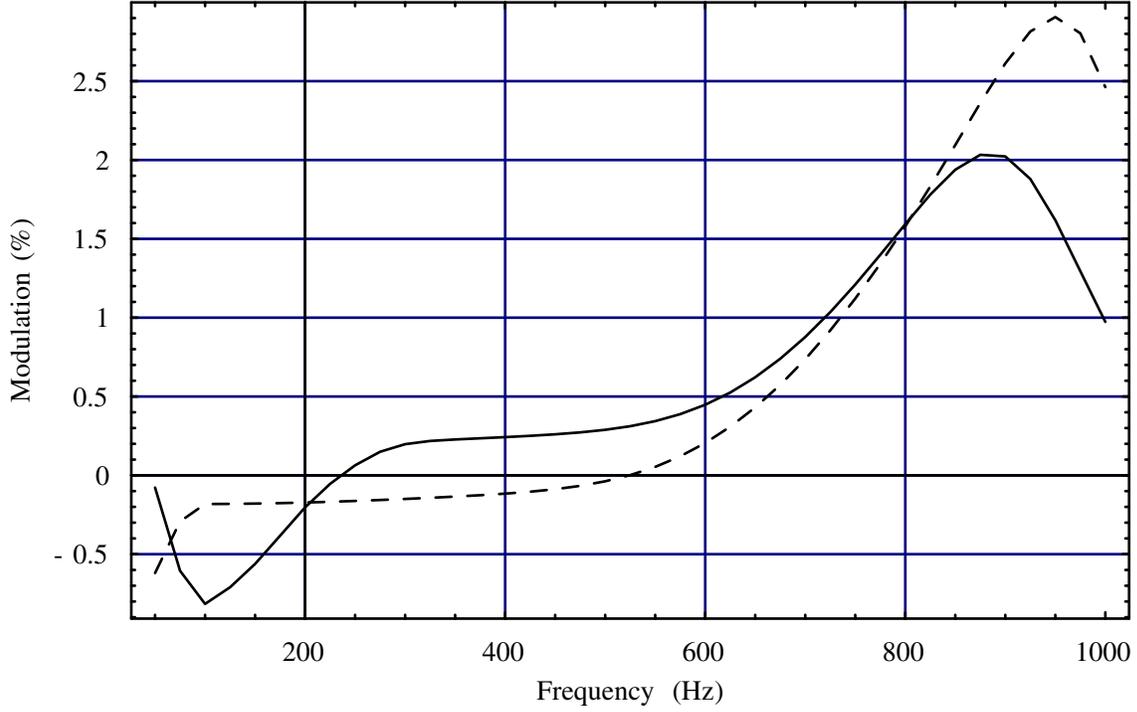}
\end{center}
\caption{\label{fig4} Amplitude of the dipole anisotropy intrinsic
modulation on the GW background as a function of the frequency.
Solid line is for GRB rate evolution, while dashed line is for
constant rate.  In this plot a negative amplitude corresponds to a
modulation in opposition of phase.}
\end{figure}
This frequency shift produces at a given frequency a diurnal
modulation of the Rayleigh power in the true signal only, which
depends from the slope of the spectrum. From the \Fref{fig2} we
can infer that this modulation will be particularly enhanced at
the extremes of the range of cosmological red shifted distribution
of the characteristic spectrum. In \Fref{fig4} we have reported
the theoretical maximum amplitude of the intrinsic modulation,
induced by the dipole anisotropy as a function of the frequency,
calculated substituting $\omega'_0$ given by Eq.\
(\ref{eq:signa1}) to $\omega_0$ into Eq.\ (\ref{eq:back4}).

A second signature of the cosmological signal comes from the fact
that the detectable power is obtained accumulating many short
duration GW pulse trains.  The signal auto correlation spectrum
defined as
\begin{equation}\label{eq:signa2}
  A(\tau)= \frac{1}{T}\int_0^T\, h(t)\,h(t+\tau)\,dt
\end{equation}
should show the evidence of a correlation over the characteristic
scale of the stellar collapse (of the order of 10 ms.) If the
instrumental noise is white and the sampling frequency of the
detector very high, the distinction could be very clear. In real
life the detection of this feature could be much harder because
the noise of the detector will be the superimposition of white
noise due to microscopic processes and colored noise coming from
other physical sources (like for example the mirror resonance or
the 1/f noise) and the sampling frequency could be inadequate.
%\begin{center}
%\includegraphics{fig5.eps}
%\end{center}
%\caption{\label{fig5} Solid line is the auto correlation function
%of the stochastic signal. Dot-dashed line is the auto correlation
%function of the colored noise in VIRGO (see text).}
%\end{figure}
\section{Conclusions}\label{sect:concl}

In \S\ref{sect:single} we have shown that the GW emission from
single bursts at cosmological distances, if the e.m. emission is
beamed with a small angle as suggested by afterglow observations,
is expected to be, for conservative values of the source
emissivity, $d_L\,\tilde h\lesssim
10^{-27}\;\mathrm{Gpc}/\sqrt{\mathrm{Hz}}$ which is well below the
detection threshold of presently planned experiments. However, as
we have discussed in \S\ref{sect:back}, if the e.m. emission is
beamed only a small fraction (one over 500) of the GRB are
observed by $\gamma$-ray satellites. This implies that small
amplitude pulse trains of GW impinge over the detector at a
frequency that could be as high as one per minute, which is
practically a continuous signal. Even if the individual pulse is
small, integrating over a reasonable observation time (order of
one year) an excess Rayleigh power should emerge from the
instrumental noise. The frequency spectrum of this eventual excess
should give a direct information on the characteristic time scale
of the collapse and on the cosmological evolution of the GRB rate
in the recent past ($z\approx 1-2$). The predicted amplitude is
conservatively estimated to be in the range of $5\times
10^{-26}\;1/\sqrt{\mathrm{Hz}}$ averaging the Rayleigh power over
one year. This estimate is rather robust because does not depends
on the beaming factor and depends only slightly from the large
scale distribution of the GRB sources. The cosmological origin of
the excess noise can be proved by detecting a dipole anisotropy,
that in the relevant frequency band could reach the level of a
fraction of percent. In addition the auto correlation spectrum of
the noise shall carry an imprint of the characteristic duration of
the GW pulse trains. The detection of those two signatures, even
if perhaps not possible with presently planned experiments, can
give the important additional evidence of the cosmological origin
of the stochastic signal and informations on the physics of GRB's.

\end{document}